\documentclass[11pt,a4paper]{article}
\usepackage{jcappub,natbib}
\usepackage{braket}
\input{colordvi.tex}
\title{Parity violation of primordial magnetic fields in the CMB bispectrum}
\author[a]{Maresuke Shiraishi}
\affiliation[a]{Department of Physics and Astrophysics, Nagoya University, 
Nagoya 464-8602, Japan} 
\emailAdd{mare@nagoya-u.jp}
\abstract{We study the parity violation in the cosmic microwave background (CMB) bispectrum induced by primordial magnetic fields (PMFs). Deriving a general formula for the CMB bispectrum generated from not only non-helical but also helical PMFs, we find that helical PMFs produce characteristic signals, which disappear in parity-conserving cases, such as the intensity-intensity-intensity bispectra arising from $\sum_{n=1}^3 \ell_n = {\rm odd}$. For fast numerical calculation of the CMB bispectrum, we reduce the one-loop formula to the tree-level one by using the so-called pole approximation. Then, we show that the magnetic anisotropic stress, which depends quadratically on non-helical and helical PMFs and acts as a source of the CMB fluctuation, produces the local-type non-Gaussianity. Comparing the CMB bispectra composed of the scalar and tensor modes with the noise spectra, we find that assuming the generation of the nearly scale-invariant non-helical and helical PMFs from the grand unification energy scale ($10^{14} {\rm GeV}$) to the electroweak one ($10^{3} {\rm GeV}$), the intensity-intensity-intensity bispectrum for $\sum_{n=1}^3 \ell_n = {\rm odd}$ can be observed by the WMAP experiment under the condition that $B_{1 \rm Mpc}^{2/3} {\cal B}_{1 \rm Mpc}^{1/3} > 2.7 - 4.5 {\rm nG}$ with $B_{1 \rm Mpc}$ and ${\cal B}_{1 \rm Mpc}$ being the non-helical and helical PMF strengths smoothed on 1 Mpc, respectively.}
\keywords{primordial magnetic fields, non-gaussianity, cosmology of theories beyond the SM, cosmological parameters from CMBR}
\arxivnumber{1202.2847}
\begin{document}
\maketitle
\flushbottom
\allowdisplaybreaks[4]

%\pacs{98.80.Cq}

%\maketitle
\def\up{\;\raise1.0pt\hbox{$'$}\hskip-6pt\partial\;}
\def\down{\;\overline{\raise1.0pt\hbox{$'$}\hskip-6pt\partial\;}}
%~~~~~~~~~~~~~~~~~~~~~~~~~~~~~~~~~~~~~~~~~~~~~~~~~~~~~~~~~~~~~~~~~~~~~~~~~~~~
%%%%%%%%%%%%%%%%%%%%%%%%%%%%%%%%%%%%%%%%%%%%%%%%%%%%%%%%%%%%%%
\section{Introduction}

The cosmological parity violation is a key feature of ultra-violet completion of general relativity and hence a lot of researchers have extracted their signals from several cosmological phenomena \cite{Lue:1998mq, Alexander:2004wk, Lyth:2005jf, Alexander:2007qe, Satoh:2007gn, Takahashi:2009wc, Barnaby:2010vf}. In particular, the effects on the cosmic microwave background (CMB) have been well-studied and the cosmological parity violation has been verified by analyzing the non-vanishing cross-correlated power spectra between the intensity ($I$) and $B$-mode polarization ($B$) anisotropies and those between $E$-mode ($E$) and $B$-mode polarization anisotropies \cite{Alexander:2006mt, Saito:2007kt, Gluscevic:2010vv, Gruppuso:2010nd, Sorbo:2011rz, Xia:2012ck}. Furthermore, beyond the linear-order effects, the impacts of the parity violation on the graviton non-Gaussianities have recently been discussed \cite{Maldacena:2011nz, Soda:2011am, Shiraishi:2011st}. According to ref.~\cite{Shiraishi:2011st}, unlike the parity-conserving non-Gaussianity, the parity-violating one induces the signals arising from $\sum_{n=1}^3 \ell_n = {\rm odd}$ in the CMB $III, IIE, IEE, EEE, IBB$ and $EBB$ bispectra and also those coming from $\sum_{n=1}^3 \ell_n = {\rm even}$ in the CMB $IIB, IEB, EEB$ and $BBB$ bispectra. In these correlations, the $III$ bispectrum from $\sum_{n=1}^3 \ell_n = {\rm odd}$ is expected to bring in the detectable information of the parity violation. 

On the other hand, if there exists the primordial magnetic field (PMF), which is a favored candidate for the seed field of microgauss-level magnetic fields in galaxies and cluster of galaxies \cite{Bernet:2008qp, Wolfe:2008nk, Kronberg:2007dy}, their power spectrum may involve the parity-violating component \cite{Field:1998hi, Anber:2006xt, Campanelli:2008kh, Campanelli:2008tt, Durrer:2010mq}. Like the above non-magnetic cases, this so-called helical PMF induces the characteristic signals in the CMB $IB$ and $EB$ correlations \cite{Caprini:2003vc, Campanelli:2004pm, Kahniashvili:2005xe, Kahniashvili:2006hy, Kunze:2011bp}. Although concrete limits on the magnitude of the helical PMF have not obtained yet, these studies imply that the $IB$ and $EB$ correlations are detectable if helical PMFs have nanogauss-level magnitudes (at the present time) and these spectra are nearly scale invariant. 
However, assuming the Gaussianity of the PMF, the beneficial signals are generated also in the CMB bispectra due to the quadratic dependence of the CMB fluctuation on the PMF. In the case where only non-helical PMFs exist, the contributions of PMFs to the primordial non-Gaussianities and the CMB bispectra have been deeply investigated in refs.~\cite{Brown:2005kr, Seshadri:2009sy, Caprini:2009vk, Cai:2010uw, Trivedi:2010gi, Shiraishi:2010yk, Shiraishi:2011dh, Shiraishi:2011fi, Shiraishi:2012rm, Barnaby:2012tk}. 

%--
In this paper, we newly consider the effects of both non-helical and helical PMFs on the CMB bispectrum. Based on our computation approach \cite{Shiraishi:2010yk, Shiraishi:2011dh, Shiraishi:2011fi, Shiraishi:2012rm}, we derive a general formula for the CMB bispectrum induced by the non-Gaussianity of the PMF anisotropic stress coming from not only non-helical PMFs but also helical ones. Then, we confirm the existence of the foregoing parity-violating signals such as the $III$ bispectrum from $\sum_{n=1}^3 \ell_n = {\rm odd}$. By the pole approximation mentioned in ref.~\cite{Shiraishi:2012rm}, we reduce this formula to a form suitable for the fast calculation in the case where the non-helical and helical PMFs have the nearly scale-invariant spectra. In this process, it is shown that the bispectrum of the PMF anisotropic stresses has the local-type shape even if helical PMFs exist. Computing the CMB $III$ bispectra composed of the scalar and tensor modes and estimating the signal-to-noise ratio, we analyze how the helical PMF affects the CMB bispectrum and show how large the PMF strength is required for the detection of the $III$ signals from $\sum_{n=1}^3 \ell_n = {\rm odd}$. 

%---
This paper is organized as follows. In the next section, we summarize the expressions and statistical properties of both non-helical and helical PMFs. In section \ref{sec:CMB_bis}, the analytic formulae and numerical results of the CMB bispectra, and the signal-to-noise ratio are presented. The final section is devoted to the summary and discussion of this paper. 
Throughout this paper, we obey the definition of the Fourier transformation as 
\begin{eqnarray}
f({\bf x}) \equiv \int \frac{d^3 {\bf k}}{(2 \pi)^3} \tilde{f}({\bf k}) e^{i {\bf k} \cdot {\bf x}}~,
\end{eqnarray}
and the rule that the subscripts and superscripts of the Greek characters and alphabets run from 0 to 3 and from 1 to 3, respectively.

%%%%%%%%%%%%%%%%%%%%%%%%%%%%%%%%%%%
\section{Statistical properties of non-helical and helical magnetic fields}\label{sec:PMF}
%%%%%%%%%%%%%%%%%%%%%%%%%%%%%%%%%%%

Let us take into account the large-scale primordial magnetic field (PMF), $B^b({\bf x},\tau)$, which is generated in the very early Universe and behaves as a source of the CMB fluctuation, on the homogeneous background and small perturbative Universe as $ds^2 = a^2 [-d\tau^2 + 2 h_{0b} d\tau dx^b + (\delta_{bc} + h_{bc}) dx^b dx^c]$. Here, $a$ and $\tau$ denote the scale factor and conformal time, respectively. Neglecting the effects of the back reaction of the fluid on the evolution of magnetic fields and considering the flux conservation, the PMF evolves as $B^b({\bf x},\tau) \propto 1/{a(\tau)}^2$. 
Each component of the energy momentum tensor composed of the PMF is given by 
\begin{eqnarray}
\begin{split}
T^0_{~0}(x^\mu) &= - \frac{1}{8 \pi a^4} B^2({\bf x}) 
%\equiv - \rho_\gamma(\tau) \Delta_B({\bf x})
~, \\
T^0_{~c}(x^\mu) &= T^b_{~0}(x^\mu) = 0~, \\
T^b_{~c}(x^\mu) &= \frac{1}{4\pi a^4} \left[\frac{B^2({\bf x})}{2}\delta^b_{~c}
 -  B^b({\bf x})B_c({\bf x})\right] 
%\\
%&\equiv \rho_\gamma(\tau) 
%\left[ \Delta_B({\bf x}) \delta^b_{~c} + \Pi^b_{Bc}({\bf x}) \right]
~. \label{eq:EMT_PMF} 
\end{split}
\end{eqnarray}
The spatial parts in Fourier space are expressed as
\begin{eqnarray}
\begin{split}
T^b_{~c}({\bf k},\tau) &\equiv {\rho_\gamma}(\tau)
\left[ \delta^b_{~c} \Delta_B({\bf k}) + \Pi_{Bc}^b({\bf k}) \right]~, \\
\Delta_B({\bf k}) &= {1 \over 8\pi \rho_{\gamma,0}}
\int \frac{d^3 {\bf k}'}{(2\pi)^3} 
B^b({\bf k'}) B_b({\bf k} - {\bf k'}) ~, \\
\Pi^b_{Bc}({\bf k}) &=-{1 \over 4\pi \rho_{\gamma,0}} \int \frac{d^3
 {\bf k'}}{(2 \pi)^3} B^b({\bf k'}) B_c({\bf k} - {\bf k'})~,
\label{eq:EMT_PMF_fourier} 
\end{split}
\end{eqnarray}
where $\rho_{\gamma} = \rho_{\gamma, 0} a^{-4}$ is the photon energy density with $\rho_{\gamma, 0}$ being its present value. 
After this, for simplicity of calculation, we ignore the trivial
time-dependence and hence the index can be lowered by 
$\delta_{bc}$. 

Conventionally, assuming the Gaussianity of the PMF, the power spectrum is expressed as \cite{Caprini:2003vc}
\begin{eqnarray}
\braket{B_a ({\bf k}) B_b ({\bf k'})} = \frac{(2 \pi)^3}{2}  
 \left[ P_B(k) P_{ab}(\hat{\bf k}) + i \eta_{abc} \hat{k}_c P_{{\cal B}}(k) \right] \delta({\bf k} + {\bf k'}) ~, \label{eq:mag_power_def}
\end{eqnarray}
where $\hat{\bf k}$ is a unit vector, $\eta_{abc}$ is the 3D Levi-Civita tensor normalized by $\eta_{123} = 1$, and $P_{ab}(\hat{\bf k}) \equiv \delta_{ab} - \hat{k}_a \hat{k}_b$ is a projection tensor coming from the divergence free nature of PMFs.  The first and second terms in the bracket represent non-helical and helical contributions, respectively. For a mathematical relation: 
\begin{eqnarray}
 \lim_{{\bf k'} \to {- \bf k}} 
 \Braket{ {\bf B}({\bf k}) \cdot {\bf B}({\bf k'}) }
\geq 
\lim_{{\bf k'} \to {- \bf k}} 
\left| \Braket{
\left[ \hat{\bf k} \times {\bf B}({\bf k}) \right] \cdot {\bf B}({\bf k'}) } \right|  \nonumber ~,
\end{eqnarray}
the power spectra of non-helical and helical PMFs obey such a magnitude relation as 
\begin{eqnarray}
P_B(k) \geq |P_{\cal B}(k)| ~. \label{eq:P_mag_rel}
\end{eqnarray}
In order to formulate the CMB bispectrum, it is convenient to use a normalized divergenceless polarization vector in two circular states, $\epsilon_a^{(\pm 1)}$, as shown in appendix \ref{appen:polarization}. Then, the above expression changes to   
\begin{eqnarray}
\braket{B_a ({\bf k}) B_b ({\bf k'})} &=& \frac{(2 \pi)^3}{2} 
\delta({\bf k} + {\bf k'}) 
\sum_{\sigma = \pm 1} 
\left[ P_B(k) - \sigma P_{{\cal B}}(k) \right] 
\epsilon_a^{(\sigma)}(\hat{\bf k}) \epsilon_b^{(-\sigma)}(\hat{\bf k}) 
\nonumber \\ 
%---
&=& \frac{(2 \pi)^3}{2} 
\delta({\bf k} + {\bf k'})
\sum_{\sigma = \pm 1} 
\left[ P_B(k') - \sigma P_{{\cal B}}(k') \right] 
\epsilon_a^{(- \sigma)}(\hat{\bf k'}) \epsilon_b^{(\sigma)}(\hat{\bf k'})
~.
\label{eq:mag_power} 
\end{eqnarray}
This implies that the second terms of the brackets in the first and second equalities creates the difference of the magnetic power spectra between two circular states as $\sigma = \pm 1$. 

To parametrize the magnetic field strengths, we introduce the quantities smoothed on $r$ as 
\begin{eqnarray}
B_{r}^2 
&\equiv& \Braket{{\bf B}({\bf x}) \cdot {\bf B}({\bf x})} |_r \nonumber \\
&=& \int \frac{d^3 {\bf k}}{(2 \pi)^3} \int \frac{d^3 {\bf k'}}{(2 \pi)^3} 
\Braket{B_a({\bf k}) B_a({\bf k'})} 
e^{-r^2 (k^2 + k'^2) / 2}  e^{i ({\bf k} + {\bf k'}) \cdot {\bf x}} ~, \\
%-----
{\cal B}_{r}^2 
&\equiv& r | \Braket{{\bf B}({\bf x}) 
 \cdot [ \nabla \times {\bf B}({\bf x}) ]  }  | |_r \nonumber \\
&=& \int \frac{d^3 {\bf k}}{(2 \pi)^3} \int \frac{d^3 {\bf k'}}{(2 \pi)^3} 
r | i k' \hat{k}_b \eta_{abc} 
\Braket{ B_a({\bf k}) B_c({\bf k'}) }  | 
e^{-r^2 (k^2 + k'^2) / 2}  e^{i ({\bf k} + {\bf k'}) \cdot {\bf x}} ~.
\end{eqnarray}
%Although the shapes of the magnetic power spectra should depend on the generation mechanism of the PMF, 
Assuming the simple power-law spectra as
\begin{eqnarray}
P_B(k) = {A}_B k^{n_B} ~, \ \ 
%------
P_{\cal B}(k) = {A}_{\cal B} k^{n_{\cal B}} ~, 
\end{eqnarray}
the spectral amplitudes are written as
\begin{eqnarray}
A_B = \frac{(2 \pi)^{n_B + 5} B^2_r}{\Gamma
 \left(\frac{n_B+3}{2}\right) k_r^{n_B + 3}} ~, \ \
%------
|A_{\cal B}| = \frac{(2 \pi)^{n_{\cal B} + 5} {\cal B}^2_r}{\Gamma
 \left(\frac{n_{\cal B} + 4}{2}\right) k_r^{n_{\cal B} + 3}} ~, 
\end{eqnarray}
where $\Gamma$ is the Gamma function and $k_r = 2\pi / r$. Note that unlike $A_B$, $A_{\cal B}$ can take both positive and negative values. 
Equation~(\ref{eq:P_mag_rel}) leads to a constraint on the PMF strengths: $|{\cal B}_r / B_r | \leq 0.288$ if the PMF spectra have nearly scale invariant shapes as $n_B = n_{\cal B} = -2.9$. 

The PMF anisotropic stress, $\Pi_{B ab}$, induces the CMB anisotropy and therefore we require their bispectrum for the computation of the CMB bispectrum. Considering equation~(\ref{eq:mag_power}), this can be straightforwardly calculated as 
\begin{eqnarray}
&& \Braket{\Pi_{B ab}({\bf k_1}) \Pi_{B cd}({\bf k_2}) \Pi_{B ef}({\bf k_3}) } 
 = (- 4\pi \rho_{\gamma,0})^{-3} 
\left[ \prod_{n=1}^3 \int d^3 {\bf k_n'} 
\sum_{\sigma_n = \pm 1}
\left\{ P_B(k_n') - \sigma_n P_{{\cal B}}(k_n') \right\}  \right] \nonumber \\
&&\qquad \times \delta({\bf k_1} - {\bf k_1'} + {\bf k_3'}) 
\delta({\bf k_2} - {\bf k_2'} + {\bf k_1'}) 
\delta({\bf k_3} - {\bf k_3'} + {\bf k_2'}) \nonumber \\
&&\qquad \times 
\frac{1}{8} 
\left[ 
\epsilon_a^{(\sigma_1)}(\hat{\bf k_1'}) \epsilon_d^{(-\sigma_1)}(\hat{\bf k_1'}) 
\epsilon_b^{(- \sigma_3)}(\hat{\bf k_3'}) \epsilon_e^{(\sigma_3)}(\hat{\bf k_3'})
\epsilon_c^{(\sigma_2)}(\hat{\bf k_2'}) \epsilon_f^{(-\sigma_2)}(\hat{\bf k_2'}) 
\right. \nonumber \\
&&\left. \qquad\qquad 
+ \{a \leftrightarrow b \ {\rm or} \ c \leftrightarrow d \ {\rm or} \ e \leftrightarrow f\} \right]~, \label{eq:ani_bis_ex}
\end{eqnarray}
where $k_D$ is the Alfv\'en-wave damping length scale
\cite{Jedamzik:1996wp, Subramanian:1997gi} as $k_D^{-1} \sim {\cal
O}(0.1)\rm Mpc$ and the curly bracket denotes the symmetric seven terms
under the permutations of indices: $a \leftrightarrow b$, $c
\leftrightarrow d$, or $e \leftrightarrow f$. For the sake of avoiding the IR divergence, we limit the range of the PMF spectral indices as $n_B, n_{\cal B} > -3$. 

Like the discussion in ref.~\cite{Shiraishi:2012rm}, if the tilts of the PMF spectra are enough red as $n_B \sim n_{\cal B} \sim -3$, the shape of the bispectrum of PMF anisotropic stresses depends strongly on the behaviors of the integrands at around three poles, namely, $k_1', k_2', k_3' \sim 0$. In this limit, the bispectrum (\ref{eq:ani_bis_ex}) reduces to 
\begin{eqnarray}
&&\Braket{\Pi_{B ab}({\bf k_1}) \Pi_{B cd}({\bf k_2}) \Pi_{B ef}({\bf k_3})}
\sim (- 4\pi \rho_{\gamma,0})^{-3} 
 \delta\left(\sum_{n=1}^3 {\bf k_n}\right) 
 \frac{\alpha A_B}{n_B + 3} k_*^{n_B + 3} \frac{8\pi}{3}\frac{1}{8}
\nonumber \\
&& \qquad %\qquad 
\times 
\left[ \sum_{\sigma_2, \sigma_3 = \pm 1}
\left\{ P_B(k_1) - \sigma_3 P_{\cal B}(k_1) \right\}
\left\{ P_B(k_2) - \sigma_2 P_{\cal B}(k_2) \right\} 
\right. \nonumber \\
%---
&&\left. \qquad\qquad\qquad\quad \times 
\delta_{a,d} 
\epsilon_b^{(\sigma_3)}(\hat{\bf k_1}) \epsilon_e^{(- \sigma_3)}(\hat{\bf k_1}) 
\epsilon_c^{(\sigma_2)}(\hat{\bf k_2}) \epsilon_f^{(- \sigma_2)}(\hat{\bf k_2}) 
 \right. \nonumber \\
%---------------
&&\qquad\quad %\qquad\quad 
\left. 
+ \sum_{\sigma_1, \sigma_2 = \pm 1}
\left\{ P_B(k_1) - \sigma_1 P_{\cal B}(k_1) \right\}
\left\{ P_B(k_3) - \sigma_2 P_{\cal B}(k_3) \right\} 
\right. \nonumber \\
%---
&&\left. \qquad\qquad\qquad\quad \times 
\epsilon_a^{(\sigma_1)}(\hat{\bf k_1}) \epsilon_d^{(- \sigma_1)}(\hat{\bf k_1}) 
\delta_{b,e} 
\epsilon_c^{(- \sigma_2)}(\hat{\bf k_3}) \epsilon_f^{(\sigma_2)}(\hat{\bf k_3}) 
 \right. \nonumber \\
%---------------
&&\qquad\quad  
\left. 
+ \sum_{\sigma_1, \sigma_3 = \pm 1} 
\left\{ P_B(k_2) - \sigma_1 P_{\cal B}(k_2) \right\}
\left\{ P_B(k_3) - \sigma_3 P_{\cal B}(k_3) \right\} 
\right. \nonumber \\
%---
&&\left. \qquad\qquad\qquad\quad \times 
\epsilon_a^{(- \sigma_1)}(\hat{\bf k_2}) \epsilon_d^{(\sigma_1)}(\hat{\bf k_2}) 
\epsilon_b^{(- \sigma_3)}(\hat{\bf k_3}) \epsilon_e^{(\sigma_3)}(\hat{\bf k_3}) 
\delta_{c,f}
 \right. \nonumber \\
%---------------
&&\qquad\quad%\qquad\quad 
\left. 
+ \{a \leftrightarrow b \ {\rm or} \ c \leftrightarrow d \ {\rm or} \ e \leftrightarrow f\} \right]~, \label{eq:ani_bis_app}
\end{eqnarray}
where $k_* \equiv 10 {\rm Mpc}^{-1}$ and $\alpha$ are a cutoff scale of the integrand and a parameter fixing the uncertainty of the amplitude associated with the approximation, respectively, and we have evaluated an integral at around each pole by following
  \begin{eqnarray}
\int d^3 {\bf k'} 
\sum_{\sigma = \pm 1}
\left\{ P_B(k') - \sigma P_{\cal B}(k') \right\} 
\epsilon_a^{(\sigma)}(\hat{\bf k'}) \epsilon_b^{(- \sigma)}(\hat{\bf k'}) 
= \frac{\alpha A_B}{n_B + 3} k_*^{n_B + 3} 
\frac{8\pi}{3}\delta_{ab} ~. \label{eq:pole_int}
\end{eqnarray}
In equation~(\ref{eq:pole_int}), due to the summation over $\sigma = \pm 1$, the contribution of the second term of the bracket vanishes. This implies that the effects of the helical PMF are tiny at around each pole. From equation~(\ref{eq:ani_bis_app}), we can see that the bispectrum of the PMF anisotropic stresses dominates at the squeezed limit such as $k_1 \approx k_2 \gg k_3$ and has the identical $k$-dependence to the local-type bispectrum of curvature perturbations \cite{Komatsu:2001rj}. Thus, we conclude that with and without helical PMFs, the shape of the non-Gaussianity associated with the PMF anisotropic stress is classified into the local-type configuration. 

In the next section, we compute the CMB bispectra generated from the non-Gaussianity of the PMF anisotropic stresses. 

%%%%%%%%%%%%%%%%%%%%%%%%%%%%%%%%%%%
\section{CMB bispectrum from non-helical and helical magnetic fields}\label{sec:CMB_bis}
%%%%%%%%%%%%%%%%%%%%%%%%%%%%%%%%%%%

In this section, we investigate the effects of both non-helical ($B$) and helical (${\cal B}$) PMFs on the CMB bispectra. At first, on the basis of the formalism presented in refs.~\cite{Shiraishi:2011fi, Shiraishi:2012rm}, we derive their exact and optimal formulae generated from equations (\ref{eq:ani_bis_ex}) and (\ref{eq:ani_bis_app}), respectively. Next, through numerical computations, we analyze the magnitudes and shapes of the CMB bispectra and examine whether the parity-violating signals coming from helical PMFs can be detected or not.

%#############################
\subsection{Formulation}
%#############################

The CMB intensity and two linear polarization fields ($X = I, E, B$) generated from the scalar-, vector- and tensor-mode perturbations ($Z = S, V, T$) are expanded by the spherical harmonics as 
\begin{eqnarray}
\frac{\Delta X^{(Z)}(\hat{\bf n})}{X} = \sum_{\ell m} a_{X, \ell m}^{(Z)} Y_{\ell m}(\hat{\bf n})~,
\end{eqnarray}
where $\hat{\bf n}$ is a line-of-sight direction. 
According to refs.~\cite{Shiraishi:2010sm, Shiraishi:2010kd}, each spherical harmonic coefficient, $a_{X, \ell m}^{(Z)}$, is given by 
\begin{eqnarray}
\begin{split}
a^{(Z)}_{X, \ell m} &= 4\pi (-i)^\ell 
\int \frac{d^3{\bf k}}{(2\pi)^3} {\cal T}_{X, \ell}^{(Z)}(k)
\sum_{\lambda} [{\rm sgn}(\lambda)]^{\lambda+x} \xi_{\ell m}^{(\lambda)}(k) ~, \\
%----------
\xi_{\ell m}^{(\lambda)}(k) &= \int d^2 \hat{\bf k} \xi^{(\lambda)}({\bf k}) 
{}_{-\lambda}Y_{\ell m}^*(\hat{\bf k}) ~,
\label{eq:alm_PMF_general}
\end{split}
\end{eqnarray}
where $\lambda = 0, \pm 1, \pm 2$ expresses the helicity of the scalar-, vector- and tensor-mode perturbations, $x = 0, 1$ discriminates the parity-even ($I, E$) and -odd ($B$) fields, and $\xi^{(\lambda)}$ and ${\cal T}_{X, \ell}^{(Z)}(k)$ are the primordial perturbation and transfer function of each mode, respectively 
\footnote{Here, we set $0^0 = 1$.}. 

If there exist large-scale PMFs, their anisotropic stresses generate additional fluctuations in the CMB. The PMF anisotropic stresses survive and become a source of the gravitational potential before neutrinos decouple and they are compensated by the neutrino anisotropic stresses. Thus, gravitational waves and curvature perturbations logarithmically grow even on superhorizon scales prior to neutrino decoupling and produce the CMB tensor- and scalar-mode anisotropies at recombination epoch \cite{Lewis:2004ef, Kojima:2009gw, Shaw:2009nf} 
\footnote{Recently, the solution of the curvature perturbation is being reanalyzed by treating the effects in both the inflationary and the radiation-dominated eras consistently \cite{Bonvin:2011dt, Bonvin:2011dr}. See however ref.~\cite{Barnaby:2012tk}.}. In contrast, due to the decaying nature of the vector potential, the CMB vector-mode anisotropies are induced by not this mechanism but the vorticities of photons enhanced by the Lorentz force from the PMF anisotropic stresses \cite{Durrer:1998ya, Mack:2001gc, Lewis:2004ef}. Consequently, we can summarize the scalar-, vector- and tensor-mode initial perturbations as 
\begin{eqnarray}
\begin{split}
\xi^{(0)}({\bf k}) &\approx 
- R_\gamma \ln\left(\frac{\tau_\nu}{\tau_B}\right) 
\frac{3}{2} O_{ab}^{(0)}(\hat{\bf k}) \Pi_{B ab}({\bf k})
%\Pi^{(0)}_{Bs}({\bf k}) 
 ~,\\
%---
\xi^{(\pm 1)}({\bf k}) &\approx 
\frac{1}{2} O_{ab}^{(\mp 1)}(\hat{\bf k}) \Pi_{B ab}({\bf k})
%\Pi^{(\pm 1)}_{Bv}({\bf k})   
~,\\
%-----
\xi^{(\pm 2)}({\bf k}) &\approx 
6 R_\gamma \ln\left(\frac{\tau_\nu}{\tau_B}\right)  
\frac{1}{2}O_{ab}^{(\mp 2)}(\hat{\bf k}) \Pi_{B ab}({\bf k})
%\Pi^{(\pm 2)}_{Bt}({\bf k}) 
~, \label{eq:initial_perturbation}
\end{split}
\end{eqnarray}
where $O_{ab}^{(0)}, O_{ab}^{(\pm 1)}$ and $O_{ab}^{(\pm 2)}$ are the projection tensors decomposing into the scalar-, vector- and tensor-mode variables, respectively, and defined in appendix~\ref{appen:polarization}.
$\xi^{(0)}$ and $\xi^{(\pm 2)}$ correspond to the curvature perturbation and gravitational wave on superhorizon scales, respectively \cite{Lewis:2004ef, Kojima:2009gw, Shaw:2009nf}, and depend on the production time of the PMF, $\tau_B$, the epoch of neutrino decoupling, $\tau_\nu \simeq 1 {\rm MeV}^{-1}$, and the ratio between the energy densities of photons and all relativistic particles, $R_\gamma \sim 0.6$, for $\tau < \tau_\nu$. As the upper and lower values of $\tau_B^{-1}$, we take the energy scales of the grand unification ($10^{14} {\rm GeV}$) and electroweak symmetry breaking ($10^3 {\rm GeV}$), corresponding to $\tau_\nu / \tau_B = 10^{17}$ and $10^6$, respectively. 
As ${\cal T}_{X, \ell}^{(S)}$ and ${\cal T}_{X, \ell}^{(T)}$, we use the standard cosmological transfer functions independent of PMFs \cite{Zaldarriaga:1996xe, Hu:1997hp, Weinberg:2008zzc, Shiraishi:2010sm} because the evolution of the cosmological perturbations are little-affected by PMFs posterior to neutrino decoupling \cite{Shaw:2009nf}. On the other hand, as ${\cal T}_{X, \ell}^{(V)}$, we should use the form including the effects of the PMF on the cosmological perturbations shown in refs.~\cite{Durrer:1998ya, Mack:2001gc, Lewis:2004ef}. 

Using equation~(\ref{eq:alm_PMF_general}), the CMB bispectrum is formulated as
\begin{eqnarray}
\Braket{\prod^3_{n=1} a_{X_n, \ell_n m_n}^{(Z_n)}}
&=& \left[ \prod^3_{n=1}
4\pi (-i)^{\ell_n}
\int_0^\infty {k_n^2 dk_n \over (2\pi)^3}
\mathcal{T}^{(Z_n)}_{X_n, \ell_n}(k_n) \sum_{\lambda_n}
 [{\rm sgn}(\lambda_n)]^{\lambda_n + x_n} \right] 
\nonumber \\
&&\times 
\Braket{\prod_{n=1}^3 \xi_{\ell_n m_n}^{(\lambda_n)}(k_n)}~. 
\label{eq:CMB_bis_PMF_general} 
\end{eqnarray}
To obtain the explicit formula for the CMB bispectrum involving the dependence on PMFs, we have to compute the angular bispectrum of the initial perturbations, $\Braket{\prod_{n=1}^3 \xi_{\ell_n m_n}^{(\lambda_n)}(k_n)}$. 
Expanding all angular dependence in the delta functions and the contractions of the projection tensors and wave number vectors in $\Braket{\prod_{n=1}^3 \xi_{\ell_n m_n}^{(\lambda_n)}(k_n)}$ by the spin spherical harmonics on the basis of appendix~\ref{appen:polarization}, and expressing the angular integrals of their spherical harmonics with the Wigner symbols \cite{Shiraishi:2011fi}, this is obtained as 
\begin{eqnarray}
\Braket{\prod_{n=1}^3 \xi_{\ell_n m_n}^{(\lambda_n)}(k_n)}
&=& \left(
  \begin{array}{ccc}
  \ell_1 & \ell_2 & \ell_3 \\
  m_1 & m_2 & m_3
  \end{array}
 \right)
(-4 \pi \rho_{\gamma,0})^{-3}
%\nonumber \\ 
%&& \times 
\sum_{L L' L''} \sum_{S, S', S'' = \pm 1} 
\left\{
  \begin{array}{ccc}
  \ell_1 & \ell_2 & \ell_3 \\
  L' & L'' & L 
  \end{array}
 \right\} 
\nonumber \\
%---
&&\times 
\left[ \prod_{n=1}^3 \int_0^{k_D} k_n'^2 dk_n' \right] 
\left\{ P_B(k_1') + S P_{\cal B}(k_1') \right\}
  \nonumber \\
%---
&&\times 
\left\{ P_B(k_2') + S' P_{\cal B}(k_2') \right\} 
\left\{ P_B(k_3') + S'' P_{\cal B}(k_3') \right\} \nonumber \\ 
%---
&&\times 
f^{S'' S \lambda_1}_{L'' L \ell_1 }(k_3',k_1',k_1) f^{S S' \lambda_2}_{L L' \ell_2}(k_1',k_2',k_2)
f^{S' S'' \lambda_3}_{L' L'' \ell_3}(k_2',k_3',k_3), \label{eq:xi_bis_pmf_SVT_exact}
\end{eqnarray}
with
%-------
\begin{eqnarray}
f^{S'' S \lambda}_{L'' L \ell}(r_3, r_2, r_1) 
&=& \sum_{L_1 L_2 L_3} \int_0^\infty y^2 dy j_{L_3}(r_3 y) j_{L_2}(r_2 y) j_{L_1}(r_1 y)  \nonumber \\
&& \times
 (-1)^{\ell + L_2 + L_3} (-1)^{\frac{L_1 + L_2 + L_3}{2}} 
I^{0~0~0}_{L_1 L_2 L_3} I^{0 S'' -S''}_{L_3 1 L''} I^{0 S -S}_{L_2 1 L} 
I_{L_1 \ell 2}^{0 \lambda -\lambda} 
 \left\{
  \begin{array}{ccc}
  L'' & L & \ell \\
  L_3 & L_2 & L_1 \\
  1 & 1 & 2
  \end{array}
 \right\} \nonumber \\
%--
&&\times
\left\{
 \begin{array}{ll}
 - \frac{2}{\sqrt{3}} (8\pi)^{3/2} R_\gamma 
\ln \left(\tau_\nu / \tau_B\right)  & (\lambda = 0) \\
 \frac{2}{3} (8\pi)^{3/2} \lambda  & (\lambda = \pm 1) \\
  - 4 (8\pi)^{3/2} R_\gamma \ln \left( \tau_\nu / \tau_B \right) &
   (\lambda = \pm 2)
 \end{array}
\right. ~. \label{eq:f_exact}
\end{eqnarray}
Here, $j_l$ is the Bessel function and the $I$ symbol is defined by
\begin{eqnarray}
I^{s_1 s_2 s_3}_{l_1 l_2 l_3}
\equiv \sqrt{\frac{(2 l_1 + 1)(2 l_2 + 1)(2 l_3 + 1)}{4 \pi}}
\left(
  \begin{array}{ccc}
  l_1 & l_2 & l_3 \\
  s_1 & s_2 & s_3
  \end{array}
 \right)~. \label{eq:I_sym}
\end{eqnarray} 
The confinement of $m_1, m_2$ and $m_3$ to the Wigner-$3j$ symbol 
$\left(
  \begin{array}{ccc}
  \ell_1 & \ell_2 & \ell_3 \\
  m_1 & m_2 & m_3
  \end{array}
 \right)$ guarantees the rotational invariance of the CMB bispectrum. 
If $A_{\cal B} = 0$, this is consistent with the corresponding equation for the non-helical case \cite{Shiraishi:2010yk, Shiraishi:2011fi, Shiraishi:2011dh, Shiraishi:2012rm}. Note that unlike $P_B$, $P_{\cal B}$ associates the spin as $S, S', S'' = \pm 1$. As described in equation~(\ref{eq:CMB_bis_PMF_numerical}), this differentiates the multipole configurations of non-helical terms from those of helical ones. Via the summations over $S, S' ,S'', \lambda_1, \lambda_2$ and $\lambda_3$, the CMB bispectrum from non-helical and helical PMFs is explicitly written as\footnote{Caution about a fact that $|\lambda|$ is determined by $Z$, namely, $|\lambda| = 0, 1, 2$ for $Z = S, V, T$, respectively.}  
\begin{eqnarray} 
\Braket{\prod^3_{n=1} a_{X_n, \ell_n m_n}^{(Z_n)}} 
&=& 
\left(
  \begin{array}{ccc}
  \ell_1 & \ell_2 & \ell_3 \\
  m_1 & m_2 & m_3
  \end{array}
 \right) 
C_{Z_1} C_{Z_2} C_{Z_3} 
\left( -4 \pi \rho_{\gamma,0} \right)^{-3} 
\sum _{L L' L''} 
\left\{
  \begin{array}{ccc}
  \ell_1 & \ell_2 & \ell_3 \\
  L' & L'' & L 
  \end{array}
 \right\}
\nonumber \\
&& \times
\sum_{\substack{L_1 L_2 L_3 \\ L'_1 L'_2 L'_3 \\ L''_1 L''_2 L''_3}} 
(-1)^{\sum_{n=1}^3\frac{L_n+L'_n+L''_n+2 \ell_n}{2}} 
I^{0~0~0}_{L_1 L_2 L_3} I^{0~0~0}_{L'_1 L'_2 L'_3} I^{0~0~0}_{L''_1
L''_2 L''_3}  \nonumber \\
&& \times
\left\{
  \begin{array}{ccc}
  L'' & L & \ell_1 \\
  L_3 & L_2 & L_1 \\
  1 & 1 & 2
  \end{array}
 \right\}
\left\{
  \begin{array}{ccc}
  L & L' & \ell_2 \\
  L'_3 & L'_2 & L'_1 \\
  1 & 1 & 2
  \end{array}
 \right\}
\left\{
  \begin{array}{ccc}
  L' & L'' & \ell_3 \\
  L''_3 & L''_2 & L''_1 \\
  1 & 1 & 2
  \end{array}
 \right\} \nonumber \\
&& \times 
\left[ \prod\limits^3_{n=1}
(-i)^{\ell_n}
\int_0^\infty {k_n^2 dk_n \over 2 \pi^2}
\mathcal{T}^{(Z_n)}_{X_n, \ell_n}(k_n)\right] \nonumber \\
&& \times 
\int_0^\infty A^2 dA j_{L_1} (k_1 A)  
\int_0^\infty B^2 dB j_{L'_1} (k_2 B)  
\int_0^\infty C^2 dC j_{L''_1} (k_3 C) \nonumber \\ 
&& \times \int_0^{k_D} k_1'^2 dk_1' 
 \left[ P_B(k_1'){\cal Q}_{L'_3, L_2, L}^{(e)} - P_{\cal B}(k_1'){\cal Q}_{L'_3, L_2, L}^{(o)} \right] 
j_{L_2} (k_1' A) j_{L'_3} (k_1' B)  \nonumber \\
&&\times  \int_0^{k_D} k_2'^2 dk_2' 
\left[ P_B(k_2'){\cal Q}_{L''_3, L'_2, L'}^{(e)} - P_{\cal B}(k_2'){\cal Q}_{L''_3, L'_2, L'}^{(o)} \right] 
j_{L'_2} (k_2'B) j_{L''_3} (k_2'C) \nonumber \\
&& \times 
\int_0^{k_D} k_3'^2 dk_3' 
\left[ P_B(k_3'){\cal Q}_{L_3, L''_2, L''}^{(e)} - P_{\cal B}(k_3'){\cal Q}_{L_3, L''_2, L''}^{(o)} \right] 
j_{L''_2} (k_3'C) j_{L_3} (k_3'A) \nonumber \\
&& \times 
8 I_{L'_3 1 L}^{0 1 -1} I_{L_2 1 L}^{0 1 -1} I_{L''_3 1 L'}^{0 1 -1}
I_{L'_2 1 L'}^{0 1 -1} I_{L_3 1 L''}^{0 1 -1} I_{L''_2 1 L''}^{0 1 -1} \nonumber \\
&& \times 
2^{3-N_S} I_{L_1~\ell_1~2}^{0 |\lambda_1| -|\lambda_1|}
 I_{L_1'~\ell_2~2}^{0 |\lambda_2| -|\lambda_2|} I_{L_1''~\ell_3~2}^{0
 |\lambda_3| -|\lambda_3|} 
{\cal D}_{L_1, \ell_1, x_1}^{(e)} {\cal D}_{L_1', \ell_2, x_2}^{(e)} {\cal D}_{L_1'', \ell_3, x_3}^{(e)} 
~, \label{eq:CMB_bis_PMF_numerical}
\end{eqnarray}
where $N_S$ is the number of the scalar modes constituting the CMB bispectrum and 
\begin{eqnarray}
C_Z \equiv
\left\{
 \begin{array}{ll}
 - \frac{2}{\sqrt{3}} (8\pi)^{3/2} R_\gamma 
\ln \left(\tau_\nu / \tau_B\right)  & (Z = S) \\
 \frac{2}{3} (8\pi)^{3/2} & (Z = V) \\
  - 4 (8\pi)^{3/2} R_\gamma \ln \left( \tau_\nu / \tau_B \right) &
   (Z = T) 
 \end{array}
\right. ~.
\end{eqnarray} 
Here, we have introduced the filter functions as 
\begin{eqnarray}
\begin{split}
{\cal Q}_{L'_3, L_2, L}^{(e)} &\equiv (\delta_{L'_3, L+1} + \delta_{L'_3,
 |L-1|}) (\delta_{L_2, L+1} + \delta_{L_2, |L-1|}) %\\
%&\quad 
+ \delta_{L'_3, L}
 \delta_{L_2, L} ~, \\
%----
{\cal Q}_{L'_3, L_2, L}^{(o)} &\equiv (\delta_{L'_3, L+1} + \delta_{L'_3,
 |L-1|}) \delta_{L_2, L} %\\
%&\quad 
+ \delta_{L'_3, L} (\delta_{L_2, L+1} + \delta_{L_2, |L-1|}) ~, \\
%----
{\cal D}_{L_1, \ell_1, x_1}^{(e)} &\equiv (\delta_{L_1, \ell_1-2} +
 \delta_{L_1, \ell_1} + \delta_{L_1, \ell_1 + 2}) \delta_{x_1,0} %\\
%&\quad 
+  (\delta_{L_1, \ell_1-1} + \delta_{L_1, \ell_1 + 1}) \delta_{x_1,1} ~, \label{eq:filter} 
\end{split}
\end{eqnarray}
which come from the above summations and selection rules of the Wigner symbols \cite{Shiraishi:2010kd} and ensure $L_3' + L_2 = {\rm even}$, $= {\rm odd}$ and $L_1 + \ell_1 + x_1 = {\rm even}$, respectively.  
Considering these filter functions and a relation derived from the selection rules as 
\begin{eqnarray}
\sum_{n=1}^3 \left( \ell_n + x_n \right) %\nonumber \\ 
%&& \quad  
+  L_2 + L_2' + L_2'' + L_3 + L_3' + L_3'' = {\rm even} ~, 
%\nonumber \\
\end{eqnarray}
we can see that 
the four terms in equation~(\ref{eq:CMB_bis_PMF_numerical}), which are composed of an even number of the helical PMF power spectra and proportional to $A_B^3$ or $A_B A_{\cal B}^2$, give the signals under the condition as 
\begin{eqnarray}
\sum_{n=1}^3 \left( \ell_n + x_n \right) = {\rm even} ~.
\end{eqnarray}
These signals can arise from the parity-even CMB bispectrum as 
\begin{eqnarray}
\Braket{\prod_{n=1}^3 \frac{\Delta X^{(Z)}(\hat{\bf n})}{X}} = 
\Braket{\prod_{n=1}^3 \frac{\Delta X^{(Z)}(- \hat{\bf n})}{X}} ~,
\end{eqnarray} 
 and are produced even in the absence of the helical PMF, namely, $A_{\cal B} = 0$ \cite{Shiraishi:2010yk, Shiraishi:2011fi, Shiraishi:2011dh, Shiraishi:2012rm}. 
On the other hand, the other four terms in equation~(\ref{eq:CMB_bis_PMF_numerical}), which are composed of an odd number of the helical PMF power spectra and proportional to $A_B^2 A_{\cal B}$ or $A_{\cal B}^3$, contribute the signals under the condition as 
\begin{eqnarray}
\sum_{n=1}^3 \left( \ell_n + x_n \right) = {\rm odd} ~.
\end{eqnarray}
These signals are due to the parity-odd CMB bispectrum as 
\begin{eqnarray}
\Braket{\prod_{n=1}^3 \frac{\Delta X^{(Z)}(\hat{\bf n})}{X}} = 
- \Braket{\prod_{n=1}^3 \frac{\Delta X^{(Z)}(- \hat{\bf n})}{X}} ~,
\end{eqnarray} 
and can not appear unless there exists the helical PMF.

As discussed in ref.~\cite{Shiraishi:2012rm}, equations (\ref{eq:xi_bis_pmf_SVT_exact}) and (\ref{eq:CMB_bis_PMF_numerical}) also have the one-loop structure in terms of multipoles, which is originated from the sexuplicate dependence on the Gaussian PMFs, and hence the numerical calculation of the CMB bispectrum requires a great deal of time. To speed up the numerical computation, we can use the optimal formula generated from equation~(\ref{eq:ani_bis_app}). In the same manner as the formulation of the exact bispectrum (\ref{eq:CMB_bis_PMF_numerical}) \cite{Shiraishi:2012rm}, we gain the approximate formula for the CMB bispectrum of the scalar and tensor modes as 
\begin{eqnarray}
%B_{X_1 X_2 X_3, \ell_1 \ell_2 \ell_3}^{{\rm app}~(Z_1 Z_2 Z_3)}(\alpha)
\Braket{\prod^3_{n=1} a_{X_n, \ell_n m_n}^{(Z_n)}}_{\rm app} 
&=& 
\left[\frac{R_\gamma \ln(\tau_\nu / \tau_B )}{4\pi \rho_{\gamma,0}} \right]^3 
\frac{\alpha A_B}{n_B + 3} k_*^{n_B + 3} \frac{8\pi}{3}  
\sum_{L_1 L_2 L_3} (-1)^{\frac{L_1 + L_2 + L_3}{2}} I_{L_1 L_2 L_3}^{0 \ 0 \ 0}
\nonumber \\
&&\times
\left( \frac{4\pi}{3} \right)^3
\sum_{L, L', L''} 
 \left\{
  \begin{array}{ccc}
  L & L' & L'' \\
  1 & 1 & 1
  \end{array}
 \right\} 
I_{L 1 1}^{0 1 -1} I_{L' 1 1}^{0 1 -1} I_{L'' 1 1}^{0 1 -1} 
\left\{
  \begin{array}{ccc}
  \ell_1 & \ell_2 & \ell_3 \\
   L_1 & L_2 & L_3 \\
   L & L' & L'' \\
  \end{array}
 \right\}
\nonumber \\
&&\times 
2^{3 - N_S}I_{\ell_1 L_1 L}^{|\lambda_1| 0 -|\lambda_1|} I_{\ell_2 L_2 L'}^{|\lambda_2| 0 -|\lambda_2|} I_{\ell_3 L_3 L''}^{|\lambda_3| 0 -|\lambda_3|} 
\nonumber \\
&&\times 
8 \int_0^\infty y^2 dy 
\left[ \prod_{n=1}^3 
 (-i)^{\ell_n} \int_0^\infty \frac{k_n^2 dk_n}{2 \pi^2} 
{\cal T}_{X_n, \ell_n}^{(Z_n)}(k_n) j_{L_n}(k_n y) \right]
 \nonumber \\
%----- 
&&\times {\cal F}^{(Z_1 Z_2 Z_3)} ~,
\label{eq:CMB_bis_PMF_app} 
\end{eqnarray}
%The function ${\cal F}^{(Z_1 Z_2 Z_3)}$ restricts the summations over multipoles of each mode as
with
\begin{eqnarray}
{\cal F}^{(SSS)}
&=& 
3  %\nonumber \\
%&&\times 
\left[
\left\{ P_B(k_1){\cal D}_{L_1, \ell_1, x_1}^{(e)} (\delta_{L,0} + \delta_{L,2}) 
- P_{\cal B}(k_1) {\cal D}_{L_1, \ell_1, x_1}^{(o)} \delta_{L,1} \right\} 
\right. \nonumber \\
&&\left. \qquad\qquad \times 
\left\{ P_B(k_2) {\cal D}_{L_2, \ell_2, x_2}^{(e)} (\delta_{L',0} + \delta_{L',2}) 
+ P_{\cal B}(k_2){\cal D}_{L_2, \ell_2, x_2}^{(o)}\delta_{L',1} \right\}
{\cal D}_{L_3, \ell_3, x_3}^{(e)} \delta_{L'', 2} 
\right. \nonumber \\
%--------
&&\quad \left. 
+ 
\left\{ P_B(k_1){\cal D}_{L_1, \ell_1, x_1}^{(e)}(\delta_{L,0} + \delta_{L,2})
 + P_{\cal B}(k_1){\cal D}_{L_1, \ell_1, x_1}^{(o)}\delta_{L,1} \right\} 
\right. \nonumber \\
&&\left. \qquad\qquad \times
{\cal D}_{L_2, \ell_2, x_2}^{(e)} \delta_{L', 2}
\left\{ P_B(k_3) {\cal D}_{L_3, \ell_3, x_3}^{(e)} (\delta_{L'',0} + \delta_{L'',2}) 
- P_{\cal B}(k_3) {\cal D}_{L_3, \ell_3, x_3}^{(o)} \delta_{L'',1}  \right\} \right. \nonumber \\
%--------
&&\quad \left. + 
{\cal D}_{L_1, \ell_1, x_1}^{(e)} \delta_{L, 2}
\left\{ P_B(k_2) {\cal D}_{L_2, \ell_2, x_2}^{(e)} (\delta_{L',0} + \delta_{L',2})
- P_{\cal B}(k_2) {\cal D}_{L_2, \ell_2, x_2}^{(o)} \delta_{L',1} \right\} 
\right. \nonumber \\
&&\left. \qquad\qquad \times
\left\{ P_B(k_3) {\cal D}_{L_3, \ell_3, x_3}^{(e)}(\delta_{L'',0} + \delta_{L'',2}) 
 + P_{\cal B}(k_3) {\cal D}_{L_3, \ell_3, x_3}^{(o)} \delta_{L'',1} \right\} 
\right] ~, \\
%##########################
{\cal F}^{(SST)}
&=&  6 \sqrt{3}
%\nonumber \\
%----------------
%&&\times 
\left[ 
 \left\{ P_B(k_1) {\cal D}_{L_1, \ell_1, x_1}^{(e)}(\delta_{L,0} + \delta_{L,2}) 
- P_{\cal B}(k_1){\cal D}_{L_1, \ell_1, x_1}^{(o)}\delta_{L,1} \right\} 
\right. \nonumber \\
&&\qquad\qquad \left. \times
\left\{ P_B(k_2) {\cal D}_{L_2, \ell_2, x_2}^{(e)}(\delta_{L',0} + \delta_{L',2}) 
+ P_{\cal B}(k_2) {\cal D}_{L_2, \ell_2, x_2}^{(o)} \delta_{L',1} \right\}
{\cal D}_{L_3, \ell_3, x_3}^{(e)} 
 \right. \nonumber \\
%--------
&&\qquad \left. + 
3 \left\{ 
\left( P_B(k_1){\cal D}_{L_1, \ell_1, x_1}^{(e)}(\delta_{L,0} + \delta_{L,2}) 
+ P_{\cal B}(k_1) {\cal D}_{L_1, \ell_1, x_1}^{(o)}\delta_{L,1} 
\right) 
{\cal D}_{L_2, \ell_2, x_2}^{(e)} \delta_{L', 2} \right. \right. \nonumber \\ 
&&\qquad\qquad \left. \left. + 
{\cal D}_{L_1, \ell_1, x_1}^{(e)}
\delta_{L, 2}
\left( P_B(k_2){\cal D}_{L_2, \ell_2, x_2}^{(e)}(\delta_{L',0} + \delta_{L',2}) 
- P_{\cal B}(k_2){\cal D}_{L_2, \ell_2, x_2}^{(o)}\delta_{L',1} \right)
\right\}
\right. \nonumber \\
&&\left. \qquad\quad \times
\left\{ P_B(k_3) {\cal D}_{L_3, \ell_3, x_3}^{(e)}
- P_{\cal B}(k_3) {\cal D}_{L_3, \ell_3, x_3}^{(o)} \right\} 
\right] \delta_{L'', 2} ~, \\
%###################
{\cal F}^{(STT)}
&=& 108
%\nonumber \\
%----
%&&\times 
\left[ 
\left\{ P_B(k_1) {\cal D}_{L_1, \ell_1, x_1}^{(e)} (\delta_{L,0} + \delta_{L,2}) 
- P_{\cal B}(k_1) {\cal D}_{L_1, \ell_1, x_1}^{(o)} \delta_{L,1} \right\} 
\right. \nonumber \\
&&\left. \qquad\qquad\quad \times 
\left( P_B(k_2){\cal D}_{L_2, \ell_2, x_2}^{(e)} - P_{\cal B}(k_2) {\cal D}_{L_2, \ell_2, x_2}^{(o)} \right) {\cal D}_{L_3, \ell_3, x_3}^{(e)}
\right. \nonumber \\
&&\left. \quad 
+ \left\{ P_B(k_1) {\cal D}_{L_1, \ell_1, x_1}^{(e)} (\delta_{L,0} + \delta_{L,2}) 
+ P_{\cal B}(k_1) {\cal D}_{L_1, \ell_1, x_1}^{(o)} \delta_{L,1} \right\}
\right. \nonumber \\
&&\left. \qquad\qquad\quad \times 
{\cal D}_{L_2, \ell_2, x_2}^{(e)} 
\left( P_B(k_3) {\cal D}_{L_3, \ell_3, x_3}^{(e)} - P_{\cal B}(k_3) {\cal D}_{L_3, \ell_3, x_3}^{(o)} \right) 
 \right. 
\nonumber \\
%---------------
&&
\quad \left. 
+ 3 \delta_{L,2} 
\left\{ P_B(k_2) {\cal D}_{L_2, \ell_2, x_2}^{(e)}  
- P_{\cal B}(k_2) {\cal D}_{L_2, \ell_2, x_2}^{(o)} \right\} 
\right. \nonumber \\
&&\left. \qquad\qquad\quad \times 
\left\{ P_B(k_3) {\cal D}_{L_3, \ell_3, x_3}^{(e)} 
- P_{\cal B}(k_3) {\cal D}_{L_3, \ell_3, x_3}^{(o)} \right\} 
%\right. \nonumber \\
%---
%&&\left. \qquad\qquad \times 
\right] \delta_{L',2} \delta_{L'',2} ~, \\
%#####################
{\cal F}^{(TTT)}
&=& 
648 \sqrt{3} 
% \nonumber \\
%&&\times 
\left[ 
\left\{ P_B(k_1) {\cal D}_{L_1, \ell_1, x_1}^{(e)} 
- P_{\cal B}(k_1) {\cal D}_{L_1, \ell_1, x_1}^{(o)} \right\} 
\right. \nonumber \\
&&\left. \qquad\qquad\qquad \times 
\left\{ P_B(k_2) {\cal D}_{L_2, \ell_2, x_2}^{(e)}
- P_{\cal B}(k_2) {\cal D}_{L_2, \ell_2, x_2}^{(o)} \right\} 
{\cal D}_{L_3, \ell_3, x_3}^{(e)}
\right. \nonumber \\
&&\qquad\quad\left. 
+ {\cal D}_{L_1, \ell_1, x_1}^{(e)} 
\left\{ P_B(k_2) {\cal D}_{L_2, \ell_2, x_2}^{(e)} 
- P_{\cal B}(k_2) {\cal D}_{L_2, \ell_2, x_2}^{(o)} \right\} 
\right. \nonumber \\
&&\left. \qquad\qquad\qquad \times
\left\{ P_{B}(k_3) {\cal D}_{L_3, \ell_3, x_3}^{(e)}
- P_{\cal B}(k_3) {\cal D}_{L_3, \ell_3, x_3}^{(o)} \right\} 
\right. \nonumber \\
&&\qquad\quad\left. + 
\left\{ P_B(k_1) {\cal D}_{L_1, \ell_1, x_1}^{(e)}
- P_{\cal B}(k_1){\cal D}_{L_1, \ell_1, x_1}^{(o)} \right\} 
{\cal D}_{L_2, \ell_2, x_2}^{(e)} 
\right. \nonumber \\
&&\left. \qquad\qquad\qquad \times
\left\{ P_B(k_3){\cal D}_{L_3, \ell_3, x_3}^{(e)} 
- P_{\cal B}(k_3) {\cal D}_{L_3, \ell_3, x_3}^{(o)} \right\} 
\right] %\nonumber \\ 
%&&\times 
\delta_{L, 2} \delta_{L', 2} \delta_{L'', 2} ~.
\end{eqnarray}
Here, ${\cal D}_{L_1, \ell_1, x_1}^{(e)}$ (defined by equation~(\ref{eq:filter})) and 
\begin{eqnarray}
{\cal D}_{L_1, \ell_1, x_1}^{(o)} 
\equiv (\delta_{L_1, \ell_1 - 1} + \delta_{L_1, \ell_1 + 1}) \delta_{x_1,0} 
+ (\delta_{L_1, \ell_1-2} + \delta_{L_1, \ell_1} + \delta_{L_1, \ell_1 + 2}) \delta_{x_1,1}~,
\end{eqnarray}
which lead to $L_1 + \ell_1 + x_1 = {\rm even}$ and $= {\rm odd}$, respectively, and the Kronecker delta functions arise from the selection rules of the Wigner symbols. Note that although the vector modes have not been considered due to their negligible signals at large scales, we can formulate them in the same way. These functions and a selection rule as $\sum_{n=1}^3 L_n = {\rm even}$ ensure that the signals of the CMB bispectra satisfying $\sum_{n=1}^3(\ell_n + x_n) = {\rm even}$ are sourced from the terms proportional to $A_B^3$ and $A_B A_{\cal B}^2$. This is consistent with the discussion in the exact formula (\ref{eq:CMB_bis_PMF_numerical}). Likewise, we can see that the signals under $\sum_{n=1}^3(\ell_n + x_n) = {\rm odd}$ are proportional to only $A_B^2 A_{\cal B}$. This means that the contribution of $A_{\cal B}^3$, which appears in the exact formula (\ref{eq:CMB_bis_PMF_numerical}), is negligible because an integral at around each pole in the bispectrum of the PMF anisotropic stresses is independent of the helical PMF
as seen in equation~(\ref{eq:pole_int}). 

Through numerical calculations, we confirmed that these optimal formulae reconstruct the shapes of the CMB bispectra based on exact formula (\ref{eq:CMB_bis_PMF_numerical}) if $\alpha$'s are identical to the values for the non-helical case \cite{Shiraishi:2012rm} as
\begin{eqnarray}
\alpha =  
\begin{cases}
 0.3350 & (Z_1 = Z_2 = Z_3 = S) \\
 0.3473 & (Z_1 = Z_2 = S, ~ Z_3 = T) \\ 
 0.3212 & (Z_1 = S,~ Z_2 = Z_3 = T) \\
 0.2991 & (Z_1 = Z_2 = Z_3 = T)
\end{cases} ~.
\end{eqnarray}

%########################### 
\subsection{Analysis} 
%###########################

Here, we show the numerical results of the CMB bispectra and signal-to-noise ratio. To calculate the CMB bispectra, we modified the Boltzmann Code for Anisotropies in the Microwave Background (CAMB) \cite{Lewis:1999bs, Lewis:2004ef} and used the Common Mathematical Library SLATEC \cite{slatec}.

\begin{figure}[t]
\begin{tabular}{cc}
\begin{minipage}[t]{0.5\hsize}
  \begin{center}
    \includegraphics[clip]{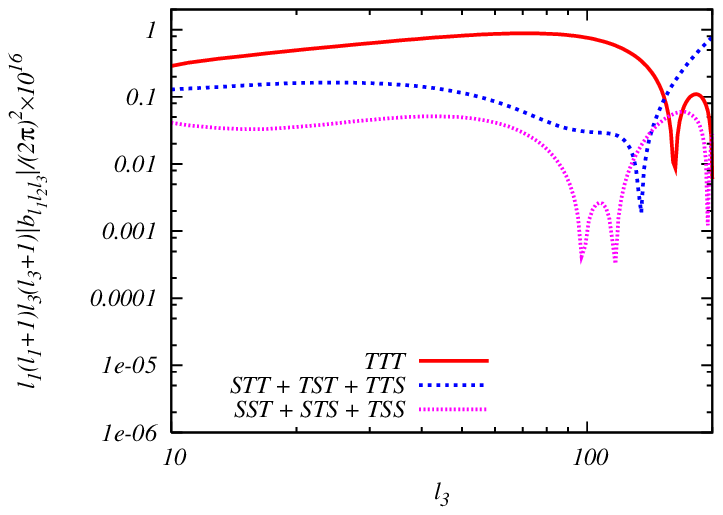}
  \end{center}
\end{minipage}
\begin{minipage}[t]{0.5\hsize}
  \begin{center}
    \includegraphics[clip]{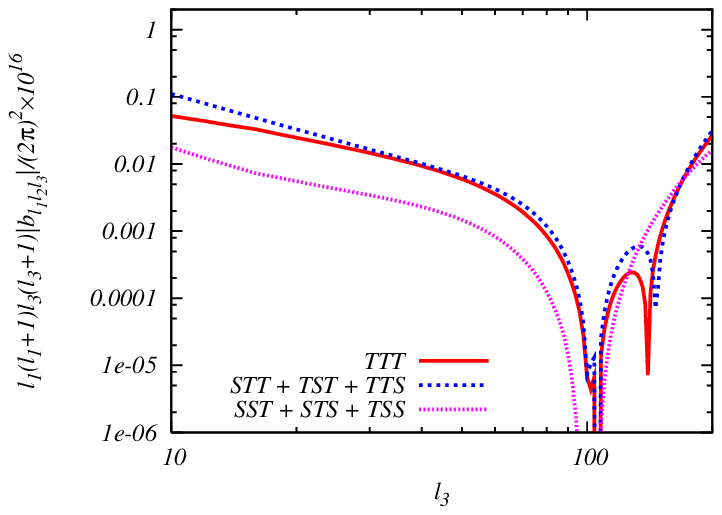}
  \end{center}
%{}caption
\end{minipage}
\end{tabular}
  \caption{Absolute values of the CMB $III$ bispectra generated from the $TTT$ (red solid line), $STT+TST+TTS$ (blue dashed one) and $SST+STS+TSS$ (magenta dotted one) modes as the function in terms of $\ell_3$ when the two multipoles are fixed as $(\ell_1, \ell_2) = (100, 105)$. The curves in the left and right panels correspond to the parity-even and -odd bispectra arising from $\sum_{n=1}^3 \ell_n = {\rm even}$ and $= {\rm odd}$, respectively.
 The PMF parameters are fixed to $B_{1 \rm Mpc} = 4.7 {\rm nG}, {\cal B}_{1 \rm Mpc} = 1.35 {\rm nG}, n_B = n_{\cal B} = -2.9$ and $\tau_\nu / \tau_B = 10^{17}$, and other parameters are identical to the mean values derived from the WMAP 7-yr data \cite{Komatsu:2010fb}}
\label{fig:bis}
\end{figure}

Figure~\ref{fig:bis} presents the parity-even signals from $\sum_{n=1}^3 \ell_n = {\rm even}$ and parity-odd ones from $\sum_{n=1}^3 \ell_n = {\rm odd}$ in the reduced intensity-intensity-intensity ($III$) bispectra of the three tensor ($TTT$), two tensor and one scalar ($STT+TST+TTS$), and one tensor and two scalar ($SST+STS+TSS$) modes defined by 
\begin{eqnarray}
  b_{III, \ell_1 \ell_2 \ell_3}^{(Z_1 Z_2 Z_3)} 
\equiv (G_{\ell_1 \ell_2 \ell_3})^{-1} 
B_{III, \ell_1 \ell_2 \ell_3}^{(Z_1 Z_2 Z_3)}  ~,
\end{eqnarray}
where  
\begin{eqnarray}
B_{X_1 X_2 X_3, \ell_1 \ell_2 \ell_3}^{(Z_1 Z_2 Z_3)} 
\equiv \sum_{m_1 m_2 m_3} 
\left(
  \begin{array}{ccc}
  \ell_1 & \ell_2 & \ell_3 \\
  m_1 & m_2 & m_3
  \end{array}
 \right) 
\Braket{\prod_{n=1}^3 a_{X_n, \ell_n m_n}^{(Z_n)}}
\end{eqnarray}
is the angle-averaged form and 
\begin{eqnarray}
G_{\ell_1 \ell_2 \ell_3} 
&\equiv& \frac{2 \sqrt{\ell_3 (\ell_3 + 1) \ell_2 (\ell_2 +
1)}}{\ell_1(\ell_1 + 1) - \ell_2 (\ell_2 + 1) - \ell_3 (\ell_3 + 1)}
\sqrt{\frac{\prod_{n=1}^3 (2 \ell_n + 1)}{4 \pi}}
\left(
  \begin{array}{ccc}
  \ell_1 & \ell_2 & \ell_3 \\
   0 & -1 & 1
  \end{array}
 \right)~.
\end{eqnarray}
The $G$ symbol is identical to $I_{\ell_1 \ell_2 \ell_3}^{0~0~0}$ when $\sum_{n=1}^3 \ell_n = {\rm even}$ \cite{Kamionkowski:2010rb, Shiraishi:2011st}. Here, we focus on the signals when multipoles have different values as $(\ell_1, \ell_2) = (100, 105)$ because the CMB bispectra from $\sum_{n=1}^3 \ell_n = {\rm odd}$ are exactly zero due to their asymmetric nature in the case where 
two of three multipoles have identical values \cite{Shiraishi:2011st}. 
Here, since we have fixed the strengths and spectral indices of the non-helical and helical PMF as $ P_B(k) = P_{\cal B}(k)$ is satisfied, it is natural prediction that the parity-even and -odd $III$ bispectra have almost same magnitudes in each mode. Actually, however, the parity-odd signals are smaller than the parity-even ones in each mode because the parity-odd signals are highly damped as three multipoles approach the similar values. The parity-even signals have same features as those in the non-helical case \cite{Shiraishi:2012rm}: the $TTT$ mode dominates for $\ell_3 \lesssim 100$ due to the Integrated Sachs Wolfe effect \cite{Pritchard:2004qp} and the bispectra including the scalar modes gradually increase for $\ell_3 \gtrsim 100$ by the acoustic oscillation. In contrast, it is hard to observe these features in the parity-odd signals due to the damping effects at $\ell_1 \sim \ell_2 \sim \ell_3 \sim 100$.

\begin{figure}[t]
\begin{tabular}{cc}
\begin{minipage}[t]{0.5\hsize}
  \begin{center}
    \includegraphics[clip]{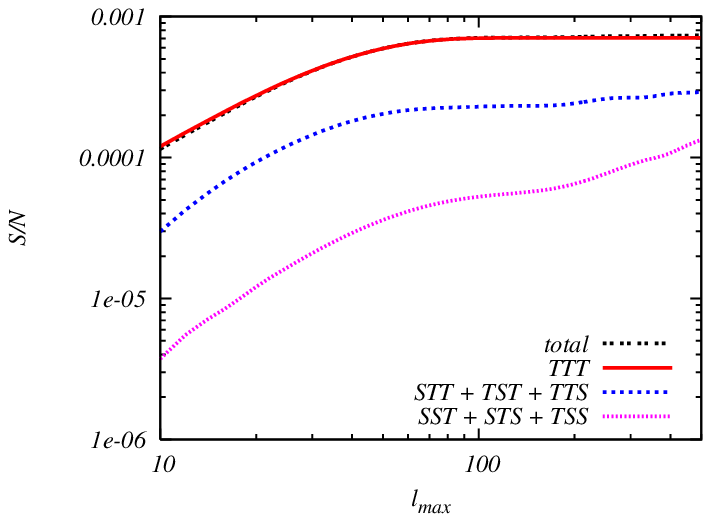}
  \end{center}
\end{minipage}
\begin{minipage}[t]{0.5\hsize}
  \begin{center}
    \includegraphics[clip]{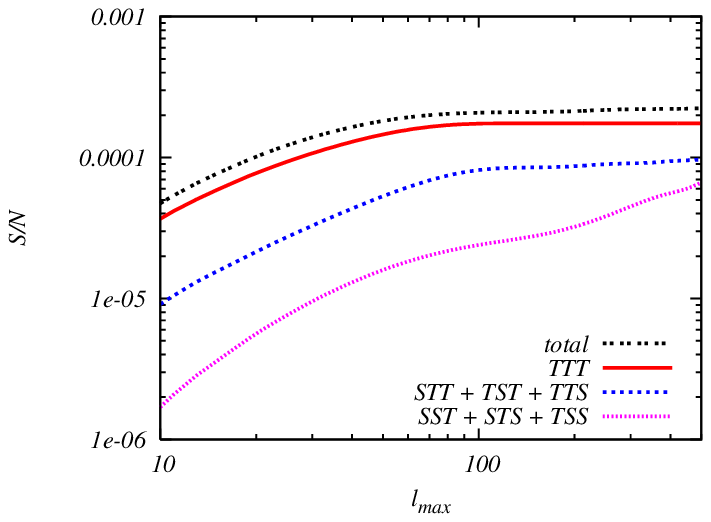}
  \end{center}
%{}caption
\end{minipage}
\end{tabular}
  \caption{Noise-free signal-to-noise ratios from the parity-even (left panel) and -odd (right one) CMB $III$ bispectra coming from $\sum_{n=1}^3 \ell_n = {\rm even}$ and $= {\rm odd}$, respectively. The ``total'' line denotes $S/N$ obtained from the total spectrum of the $TTT, STT, TST, TTS, SST, STS$ and $TSS$ modes, and the others correspond to $S/N$'s coming from each mode. Here, we fix the PMF parameters as $B_{\rm 1Mpc} = 1.0 {\rm nG}, {\cal B}_{\rm 1Mpc} = 0.287 {\rm nG}, n_B = n_{\cal B} = -2.9$ and $\tau_\nu / \tau_B = 10^{17}$. The other parameters are identical to the mean values obtained from the WMAP-7yr data \cite{Komatsu:2010fb}.}
\label{fig:SN}
\end{figure}

Figure~\ref{fig:SN} shows the signal-to-noise ratios from the parity-even and -odd CMB $III$ bispectra of the $TTT$, $STT+TST+TTS$ and $SST+STS+TSS$ modes, and their total bispectrum, respectively
\footnote{We neglect the three scalar mode for its weak signals at $\ell < 500$.}, given by 
\begin{eqnarray} 
\left(\frac{S}{N}\right)^2 =  \sum_{2 \leq \ell_1 \leq \ell_2 \leq \ell_3 \leq \ell_{\rm max}} 
\frac{\left( \sum_{Z_1 Z_2 Z_3} B_{III, \ell_1 \ell_2 \ell_3}^{(Z_1 Z_2 Z_3)} \right)^2}
{ \sigma_{\ell_1 \ell_2 \ell_3}^2 }~, \label{eq:SN_bis}
\end{eqnarray}
where $\sigma_{\ell_1 \ell_2 \ell_3}^2$ denotes the variance of the bispectrum. Assuming the weakly non-Gaussianity, this is calculated as
\begin{eqnarray}
\sigma_{\ell_1 \ell_2 \ell_3}^2
&=& \sum_{\substack{m_1 m_2 m_3 \\ m_1' m_2' m_3'}} 
\left(
  \begin{array}{ccc}
  \ell_1 & \ell_2 & \ell_3 \\
  m_1 & m_2 & m_3
  \end{array}
 \right) 
\left(
  \begin{array}{ccc}
  \ell_1 & \ell_2 & \ell_3 \\
  m_1' & m_2' & m_3'
  \end{array}
 \right) 
\Braket{\prod_{n=1}^3 a_{\ell_n m_n}^{\rm fid} a_{\ell_n m_n'}^{\rm fid}} \nonumber \\
&=& C_{\ell_1}^{\rm fid} C_{\ell_2}^{\rm fid} C_{\ell_3}^{\rm fid}  
\left[ (-1)^{\ell_1 + \ell_2 + \ell_3} (1 + 2 \delta_{\ell_1, \ell_2} \delta_{\ell_2, \ell_3}) + \delta_{\ell_1, \ell_2} + \delta_{\ell_2, \ell_3} + \delta_{\ell_3, \ell_1} \right]
~,
\end{eqnarray}
where $a_{\ell m}^{\rm fid}$ and $C_{\ell}^{\rm fid}$ are the fiducial $a_{\ell m}$ and power spectrum consistent with the current observational data. When $\sum_{n=1}^3 \ell_n = {\rm even}$, this is identical to that derived in refs.~\cite{Komatsu:2001rj, Bartolo:2004if}. 
Here, we have neglected the noise other than the cosmic variance. We can find that both in the parity-even and odd bispectra, the $TTT$ mode dominates over the signals. However, the scalar mode increasingly affects at intermediate scales, where the tensor mode reaches a plateau, and will become important at smaller scales. Although we also have taken the PMF parameters keeping $P_B(k) = P_{\cal B}(k)$, the parity-odd signals are small compared with the parity-even ones. Nevertheless, these signals are precious information to probe the parity violation of PMFs because these should vanish if ${\cal B}_{1 \rm Mpc} = 0$ at variance with the parity-even signals. As discussed in the previous subsection, the CMB $III$ bispectra ($x_1 = x_2 = x_3 = 0$) of the scalar and tensor modes from $\sum_{n=1}^3 \ell_n = {\rm odd}$ are asymptotically proportional to $B_{1 \rm Mpc}^4 {\cal B}_{1 \rm Mpc}^2 [\ln (\tau_\nu / \tau_B)]^3$. Therefore, from the value of the total signal-to-noise ratio for $\ell_{\rm max} = 500$ described in figure~\ref{fig:SN} as $S/N = 2.236 \times 10^{-4}$, we can understand that if $n_B = n_{\cal B} = -2.9$, $\tau_\nu / \tau_B = 10^{17} (10^6)$ and $B_{1 \rm Mpc}^{2/3} {\cal B}_{1 \rm Mpc}^{1/3} > 2.7 (4.5) {\rm nG}$, $S/N$ exceeds unity. 
Since $\ell_{\rm max} = 500$ roughly corresponds to the WMAP-level resolution \cite{Komatsu:2001rj}, these conditions are requirements for the detection of the helical PMF by the WMAP experiment. Due to the small enhancement of the signal-to-noise ratio at small scales by the scalar modes, we expect that these conditions will improve slightly by the use of the PLANCK data \cite{:2006uk} like the non-helical case \cite{Shiraishi:2012rm}.
%Note that due to the plateau of the signal-to-noise ratio for $\ell_{\rm max} \lesssim 1000$, we expect that these conditions will not change much even in the PLANCK experiment.
%}

%%%%%%%%%%%%%%%%%%%%%%%%%%%%%%%%%%%%%%%%%%%% 
\section{Summary and discussion}
\label{sec:summary}
%%%%%%%%%%%%%%%%%%%%%%%%%%%%%%%%%%%%%%%%%%%%

In this paper, we investigated the effects of the parity-violating helical PMF in the CMB  bispectrum. At first, following the definition of the helical PMF in ref.~\cite{Caprini:2003vc}, we calculated the bispectrum of the PMF anisotropic stresses involving both non-helical and helical PMFs. Then, through the application of the pole approximation \cite{Shiraishi:2012rm}, we found that the non-Gaussianity of the PMF anisotropic stresses closes to the local-type one if the non-helical and helical PMF power spectra have nearly scale-invariant shapes. In the same manner as our formalism \cite{Shiraishi:2011fi, Shiraishi:2012rm}, we formulated the CMB bispectrum generated from not only non-helical but also helical PMFs. Our formulae showed that the finite signals arise from not only $\sum_{n=1}^3 \ell_n = {\rm even}$ but also $\sum_{n=1}^3 \ell_n = {\rm odd}$ in the intensity-intensity-intensity bispectrum. The latter signals, which is asymptotically proportional to $B_{\rm 1Mpc}^4 {\cal B}_{\rm 1Mpc}^2$, never appear unless the helical PMF exists and hence these are good observables to probe the parity-violating information of the PMF. 
Through the numerical computation of the CMB intensity-intensity-intensity bispectra of the scalar and tensor modes and their signal-to-noise ratios, we found that the parity-odd signals from $\sum_{n=1}^3 \ell_n = {\rm odd}$ are smaller than parity-even ones from $\sum_{n=1}^3 \ell_n = {\rm even}$ since the parity-odd signals are highly suppressed in the case where $\ell_1 \sim \ell_2 \sim \ell_3$ due to the asymmetric nature of the CMB bispectrum. Nevertheless, the computation of the signal-to-noise ratio provided a fact that if $B_{1 \rm Mpc}^{2/3} {\cal B}_{1 \rm Mpc}^{1/3} > 2.7 - 4.5 {\rm nG}$ is satisfied, the parity-odd signals dominate the cosmic variance for $\ell < 500$ and it is possible to access the parity-violating nature of the PMF by the WMAP experiment. Furthermore, the information of the polarizations is expected to improve this detectability. 

This study with ref.~\cite{Shiraishi:2011st} gives a motivation to constrain the parity-violating non-Gaussianities from the observational data by using the signals under $\sum_{n=1}^3 \ell_n = {\rm odd}$ in the CMB intensity-intensity-intensity bispectrum. In ref.~\cite{Shiraishi:2011st}, we have shown that the parity-violating Weyl cubic term produces not the local-type non-Gaussianities but the equilateral-type ones in the graviton sector and they also induce the CMB intensity-intensity-intensity bispectrum for $\sum_{n=1}^3 \ell_n = {\rm odd}$. Therefore, it will become important to constrain the magnitude of each-type non-Gaussianity like $f_{\rm NL}^{\rm local}, f_{\rm NL}^{\rm equil}$ and $f_{\rm NL}^{\rm orthog}$ in the parity-even case \cite{Smith:2009jr, Komatsu:2010fb}, and differentiate between various non-Gaussian sources involving the parity violation. 
These considerations remain as future issues.

%####################################
%####################################
\acknowledgments
%------
We would like to thank to Shuichiro Yokoyama, Daisuke Nitta, and Kiyotomo Ichiki for helpful advice. This work was supported in part by a Grant-in-Aid for JSPS Research under Grant No.~22-7477, Grant-in-Aid for Scientific Research on Priority Areas No.~467 ``Probing the Dark Energy through an Extremely Wide and Deep Survey with Subaru Telescope'', and Grant-in-Aid for Nagoya University Global COE Program ``Quest for Fundamental Principles in the Universe: from Particles to the Solar System and the Cosmos'', from the Ministry of Education, Culture, Sports, Science and Technology of Japan. 
We also acknowledge the Kobayashi-Maskawa Institute for the Origin of Particles and the Universe, Nagoya University, for providing computing resources useful in conducting the research reported in this paper. 

%%%%%%%%%%%%%%%%%%%%%%%%%%%%%%%%%%%
\appendix
%%%%%%%%%%%%%%%%%%%%%%%%%%%%%%%%%%%
\section{Projection tensors}\label{appen:polarization}

In this section, we present some useful relations of the projection tensors based on refs.~\cite{Shiraishi:2010kd, Shiraishi:2012rm}. These are utilized as tools for the formulation of the CMB bispectra (\ref{eq:CMB_bis_PMF_numerical}) and (\ref{eq:CMB_bis_PMF_app}). 

An arbitrary unit vector, a normalized divergenceless vector and transverse-traceless tensor can be defined by using the spin spherical harmonics as 
\begin{eqnarray}
\begin{split}
\hat{k}_a &= \sum_m \alpha_a^{m} Y_{1 m}(\hat{\bf k}) ~, \\
%----
\epsilon_a^{(\pm 1)} (\hat{\bf k}) 
&= \mp \sum_m \alpha_a^m {}_{\pm 1} Y_{1 m} (\hat{\bf k})~, \\
%----
e_{ab}^{(\pm 2)} (\hat{\bf k}) 
&=\sqrt{2} \epsilon_a^{(\pm 1)} (\hat{\bf k}) \epsilon_b^{(\pm 1)} (\hat{\bf k}) ~,
\end{split}
\end{eqnarray}
with
\begin{eqnarray}
%----
\alpha_a^m \equiv \sqrt{\frac{2 \pi}{3}}
 \left(
  \begin{array}{ccc}
   -m (\delta_{m,1} + \delta_{m,-1}) \\
   i~ (\delta_{m,1} + \delta_{m,-1}) \\
   \sqrt{2} \delta_{m,0}
  \end{array}
\right)~,
%\end{split}
\end{eqnarray}
whose scalar products are given by 
\begin{eqnarray}
\alpha_a^m \alpha_a^{m'} = \frac{4 \pi}{3} (-1)^m \delta_{m,-m'}~, \ \
\alpha_a^m \alpha_a^{m' *} = \frac{4 \pi}{3} \delta_{m,m'}~.
\end{eqnarray}
Then, the divergenceless vector and transverse-traceless tensor obey
\begin{eqnarray}
\begin{split}
\hat{k}^a \epsilon_a^{(\pm 1)}(\hat{\bf k}) &= 0~, \\
\epsilon^{(\pm 1) *}_a (\hat{\bf k}) &= \epsilon^{(\mp 1)}_a (\hat{\bf k})
 = \epsilon^{(\pm 1)}_a (-\hat{\bf k})~, \\
%---
\epsilon^{(\lambda)}_a (\hat{\bf k}) \epsilon^{(\lambda')}_a (\hat{\bf k}) 
&= \delta_{\lambda, -\lambda'} \ \ \ ({\rm for} \ \lambda, \lambda' = \pm 1) ~, \\
%-----------
e_{aa}^{(\pm 2)}(\hat{\bf k}) &= \hat{k}_a e_{ab}^{(\pm 2)}(\hat{\bf k}) = 0~, \\
e_{ab}^{(\pm 2) *}(\hat{\bf k}) &= e_{ab}^{(\mp 2)}(\hat{\bf k}) = e_{ab}^{(\pm
2)}(- \hat{\bf k})~, \\
e_{ab}^{(\lambda)}(\hat{\bf k}) e_{ab}^{(\lambda')}(\hat{\bf k}) &= 2
\delta_{\lambda, -\lambda'} \ \ \ ({\rm for} \ \lambda, \lambda' = \pm 2)~. \label{eq:pol_tens_relation}
\end{split}
\end{eqnarray}

Using the divergenceless vector, a projection tensor in equation~(\ref{eq:mag_power_def}) and a unit vector are expressed as
\begin{eqnarray}
\begin{split}
P_{ab}(\hat{\bf k}) &= \sum_{\sigma = \pm 1} 
\epsilon_a^{(\sigma)}(\hat{\bf k}) \epsilon_b^{(-\sigma)}(\hat{\bf k})~, \\
%---
\hat{k}_c &= i \eta_{abc} \epsilon_a^{(+1)}(\hat{\bf k}) 
\epsilon_b^{(-1)}(\hat{\bf k}) ~.
\end{split}
\end{eqnarray}
These relations lead to equation~(\ref{eq:mag_power})

The projection tensors, which are expanded by the spin spherical harmonics as
\begin{eqnarray}
\begin{split}
O_{ab}^{(0)}(\hat{\bf k}) &\equiv - \hat{k}_a \hat{k}_b + \frac{1}{3} \delta_{ab} \\
&= - 2 I_{2 1 1}^{0 1 -1} \sum_{M m_a m_b} 
Y_{2 M}^*(\hat{\bf k}) \alpha_a^{m_a} \alpha_b^{m_b} 
\left(
  \begin{array}{ccc}
  2 & 1 & 1 \\
  M & m_a & m_b 
  \end{array}
 \right) ~, \\
%-----
O_{ab}^{(\pm 1)}(\hat{\bf k}) 
&\equiv 
\hat{k}_a {\epsilon}_b^{(\pm 1)}(\hat{\bf k}) 
+ \hat{k}_b {\epsilon}_a^{(\pm 1)}(\hat{\bf k}) \\
&= \pm 2\sqrt{3} I_{2 1 1}^{0 1 -1} \sum_{M m_a m_b} 
{}_{\mp 1} Y^*_{2 M}(\hat{\bf k}) \alpha_a^{m_a} \alpha_b^{m_b} 
\left(
  \begin{array}{ccc}
  2 & 1 & 1 \\
  M & m_a & m_b
  \end{array}
 \right)
~, \\
%================
O_{ab}^{(\pm 2)}(\hat{\bf k}) &\equiv 
e_{ab}^{(\pm 2)}(\hat{\bf k})
 \\
&= 2\sqrt{3} I_{2 1 1}^{0 1 -1}
\sum_{M m_a m_b} {}_{\mp 2}Y_{2 M}^*(\hat{\bf k})
\alpha_a^{m_a} \alpha_b^{m_b}
 \left(
  \begin{array}{ccc}
  2 & 1 & 1 \\
  M & m_a & m_b
  \end{array}
 \right) ~, 
\end{split} 
\end{eqnarray}
decompose an arbitrary physical tensor such as the metric or the PMF anisotropic stress into the two scalar ($\chi_{\rm iso}, \chi^{(0)}$), two vector ($\chi^{(\pm 1)}$) and two tensor ($\chi^{(\pm 2)}$) components:
\begin{eqnarray}
\chi_{ab}({\bf k}) &=& - \frac{1}{3} \chi_{\rm iso}({\bf k}) \delta_{ab} 
+ \chi^{(0)}({\bf k}) O^{(0)}_{ab}(\hat{\bf k}) \nonumber \\
&&+ \sum_{\lambda = \pm 1} \chi^{(\lambda)}({\bf k}) O^{(\lambda)}_{ab}(\hat{\bf k})
+ \sum_{\lambda = \pm 2} \chi^{(\lambda)}({\bf k}) O^{(\lambda)}_{ab}(\hat{\bf k})
~.
\end{eqnarray}
Here, $I_{2 1 1}^{0 1 -1} = \sqrt{3/(8\pi)}$ is given by equation~(\ref{eq:I_sym}). 
Considering equation~(\ref{eq:pol_tens_relation}), we can derive the inverse formulae as 
\begin{eqnarray}
\begin{split}
\chi^{(0)}({\bf k}) &= \frac{3}{2}O^{(0)}_{ab}(\hat{\bf k}) \chi_{ab}({\bf k}) ~, \\
%----
\chi^{(\pm 1)}({\bf k}) &= \frac{1}{2} O^{(\mp 1)}_{ab}(\hat{\bf k}) \chi_{ab}({\bf k})~, \\ 
%----
\chi^{(\pm 2)}({\bf k}) &= \frac{1}{2} O_{ab}^{(\mp 2)}(\hat{\bf k}) \chi_{ab}({\bf k})~. \label{eq:tensor_inverse}
\end{split}
\end{eqnarray}
These are used in the calculation of the each-mode initial perturbation (\ref{eq:initial_perturbation}). 

Finally, we describe other useful expressions for computation of the CMB approximate bispectrum (\ref{eq:CMB_bis_PMF_app}) as 
\begin{eqnarray}
\begin{split}
O_{ab}^{(0)}(\hat{\bf k})\epsilon_b^{(\sigma)}(\hat{\bf k})
&= \frac{1}{3} \epsilon_a^{(\sigma)} (\hat{\bf k}) ~, \\
%-----
e_{ab}^{(\lambda)}(\hat{\bf k})\epsilon_b^{(\sigma)}(\hat{\bf k})
&= \sqrt{2} \epsilon_a^{(\frac{\lambda}{2})}(\hat{\bf k}) \delta_{\sigma, -\frac{\lambda}{2}} ~, \\
%----
\epsilon_a^{(\sigma)}(\hat{\bf k}) \epsilon_b^{(-\sigma)}(\hat{\bf k}) 
&= - \sum_{L} I_{L 1 1}^{0 -\sigma \sigma}  \sum_{M m_a m_b} Y_{L M}^*(\hat{\bf k}) 
\alpha_a^{m_a} \alpha_b^{m_b} 
\left(
  \begin{array}{ccc}
  L & 1 & 1 \\
  M & m_a & m_b
  \end{array}
\right)~.
\end{split}
\end{eqnarray}

\bibliography{paper}
%\nocite{*}
\end{document}